\documentclass[fleqn]{jaes}

\jyear{2024}
\jmonth{Dec.}

\usepackage{amsmath}
\usepackage{amsmath,graphicx,amssymb,bm,bbm}
\usepackage{bm}
\usepackage{algorithm}
\usepackage{algpseudocodex}
\usepackage{layout}
\usepackage{subfig}
\usepackage{graphicx}     %
\usepackage{enumitem}

\usepackage{graphicx} 
\usepackage{hyperref}
\usepackage{booktabs}
\usepackage{lipsum}
\usepackage{svg}
\usepackage{adjustbox}
\usepackage{draftwatermark}
\usepackage{printlen}
\usepackage{pgfplots}
\pgfplotsset{compat=1.17}
\SetWatermarkText{}
\SetWatermarkColor[gray]{0.9}
\SetWatermarkScale{1.4}

\usepackage[nolist]{acronym}

\begin{acronym}
\acro{stft}[STFT]{Short-Time Fourier Transform}
\acro{fft}[FFT]{Fast Fourier Transform}
\acro{istft}[iSTFT]{Inverse Short-Time Fourier Transform}
\acro{dnn}[DNN]{Deep Neural Network}
\acro{mlp}[MLP]{Multi-Layer Perceptron}
\acro{pesq}[PESQ]{Perceptual Evaluation of Speech Quality}
\acro{psd}[PSD]{Power Spectral Density}
\acro{rir}[RIR]{Room Impulse Response}
\acro{snr}[SNR]{Signal-to-Noise Ratio}
\acro{lstm}[LSTM]{Long Short-Term Memory}
\acro{polqa}[POLQA]{Perceptual Objective Listening Quality Analysis}
\acro{sdr}[SDR]{Signal-to-Distortion Ratio}
\acro{estoi}[ESTOI]{Extended Short-Term Objective Intelligibility}
\acro{tf}[T-F]{Time-Frequency}
\acro{lsd}[LSD]{Log Spectral Distance}
\acro{sisdr}[SI-SDR]{Scale-Invariant Signal-to-Distortion Ratio}
\acro{mos}[MOS]{Mean Opinion Score}
\acro{clap}[CLAP]{Contrastive Audio Language Pretraining}
\acro{dps}[DPS]{Diffusion Posterior Sampling}
\acro{ode}[ODE]{Ordinary Differential Equation}
\acro{pfode}[PF-ODE]{Probability Flow Ordinary Differential Equation}
\acro{ccr}[CCR]{Cubic Catmull-Rom}
\acro{dlsd}[dLSD]{Degraded Log Spectral Distance}
\acro{fad}[FAD]{Fréchet Audio Distance}
\acro{hwr}[HWR]{Half-Wave Rectifier}
\acro{rrmse}[RR-MSE]{Ramp Response Mean Squared Error}

\end{acronym}

\begin{document}

\markboth{ŠVENTO ET AL.}{UNKOWN NONLINEAR DISTORTION}

\title{
Estimation and Restoration of Unknown Nonlinear Distortion using Diffusion }

\authorgroup{
\author{MICHAL ŠVENTO,\textsuperscript{1,2}}
\author{ELOI MOLINER,\textsuperscript{1}}
\author{LAURI JUVELA,\textsuperscript{1}}
\author{ALEC WRIGHT\textsuperscript{3}} \\
AND \author{VESA VÄLIMÄKI,\textsuperscript{1}}\role{AES Fellow}
\affil{\textsuperscript{1}Acoustics Lab, Department of Information and Communications Engineering, Aalto University, Espoo, Finland \\
\textsuperscript{2}
Dept. of Telecommunications, FEEC,
Brno University of Technology, Brno, Czech Republic \\
\textsuperscript{3}
Acoustics and Audio Group,
University of Edinburgh,
Edinburgh, UK}
}

\abstract{The restoration of nonlinearly distorted audio signals, alongside the identification of the applied memoryless nonlinear operation, is studied.
The paper focuses on the difficult but practically important case in which both the nonlinearity and the original input signal are unknown. 
The proposed method uses a generative diffusion model trained unconditionally on guitar or speech signals to jointly model and invert the nonlinear system at inference time. Both the memoryless nonlinear function model and the restored audio signal are obtained as output. 
Successful example case studies are presented including inversion of hard and soft clipping, digital quantization, half-wave rectification, and wavefolding nonlinearities. Our results suggest that, out of the nonlinear functions tested here, the cubic Catmull-Rom spline is best suited to approximating these nonlinearities. In the case of guitar recordings, comparisons with informed and supervised methods show that the proposed blind method is at least as good as they are in terms of objective metrics. Experiments on distorted speech show that the proposed blind method outperforms general-purpose speech enhancement techniques and restores the original voice quality. The proposed method can be applied to audio effects modeling, restoration of music and speech recordings, and characterization of analog recording media.
}
\maketitle

\section{INTRODUCTION}

Nonlinear distortions are a common artifact in audio signals, arising in various scenarios such as hardware limitations, intentional effects in music production, or environmental interactions during signal capture. The restoration of such distorted signals, alongside the estimation of the nonlinear function that caused the distortion, is a challenging yet crucial task and the topic of this study.

A well known example is clipping, a phenomenon that occurs when the amplitude of the audio signal exceeds the maximum limit, causing saturation \cite{zaviska2021survey}. In the case of hard-clipping, the nonlinearity is easy to characterize by analyzing the distorted measurement, and the known operator can be used to solve the declipping problem \cite{abel1991, siedenburg2014, zaviska2021survey}. However, there is an infinite range of other nonlinear distortions that can occur, and estimating the operator in those cases is often impossible without access to a clean reference signal.
The identification and modeling of nonlinear systems can be useful not only from a restoration perspective but also for tasks such as audio effects modeling \cite{eichas2016black} and characterizing or modeling analog recording media \cite{valimaki2008digital, mikkonen2023neural}.

Traditionally, characterizing nonlinear systems using data-driven methods necessitated pairs of aligned input and output audio recordings. This includes Wiener-Hammerstein models \cite{eichas2016black}, Volterra series \cite{tronchin2013emulation, bouvier2017nonlinear, orcioni2018identification, bennett2022antialiasing}, Chebyshev models \cite{novak2010chebyshev}, neural networks \cite{parker2019modelling, wright2020real, vanhatalo2024evaluation}, or other specialized signal processing pipelines \cite{yeh2024ddsp}. 
However, in many practical cases such paired data is often unavailable.
The challenge of estimating the nonlinearity based solely on the observed distorted output without paired input data is called a blind estimation problem.

Recent studies have addressed this or similar tasks using different approaches. For example, Wright et al.~proposed using adversarial training to model guitar distortion effects with unpaired data \cite{wright2023adversarial}.
Diffusion bridges have also been explored for blind effects transformation \cite{moliner2024gaussian}.
Alternatively, when a sufficiently large pre-training database is available, it has been shown recently that feature representations capturing relevant audio effect information can be learned \cite{chen2024towards, wright2024open, steinmetz2024st}. These representations have proven useful for modeling unseen effects, although their success depends heavily on the quality and diversity of the pre-training data.

Various strategies for distortion or audio effects inversion, or restoration, have been explored, ranging from sparsity-based approaches, such as A-SPADE for declipping \cite{kitic2015sparsity}, to compensating nonlinear function estimation for soft declipping \cite{avila2017audio}.
Recently, neural network-based methods have become the focus, including supervised learning  for guitar distortion removal and speech declipping \cite{Imort2022distortion, yi2024ddd}, vocoder-based methods for distortion removal and speech restoration \cite{lee2024distortion, liu2022voicefixer}, and compositional approaches for the inversion of audio effects \cite{rice2023general}.

A strategy combining estimation and restoration is the approach by Hinrichs et al. \cite{hinrichs2024blind}. Their method determinse the nonlinear system by using a blind estimate of the clean signal, obtained via a supervised restoration model, rather than the clean input itself. However, the method heavily relies on the accuracy of the restoration model. While effective in some cases, it has limitations that can hinder system identification. A major limitation is the need for paired input-output data during training. Although these pairs can be synthetically generated through randomized distortions, this can lead to generalization issues, especially with unseen distortions. Furthermore, due to the supervised framework, the model often struggles with ill-conditioned distortions, causing oversmoothing or regression to the mean. Errors in the restoration process can also propagate, degrading performance in the system identification stage.

A promising alternative to address these limitations is using generative models that jointly perform restoration and distortion estimation. 
Recent research into diffusion models for inverse problems shows promise in solving both linear and nonlinear problems when the distortion operator is known \cite{moliner2022solving, hernandez2024vrdmg, levy2023controllable}.
Notably, the framework used in these works, popularly known as \ac{dps} \cite{chung2022diffusion}, only requires training a diffusion model on clean audio files.  
Previous works have adapted this framework to blind linear inverse problems, where a linear filter or room impulse response is iteratively estimated alongside the restored audio or speech signal \cite{moliner2023zeroshot, lemercier2024buddyjournal}. While these prior studies have addressed similar challenges, none has effectively tackled blind nonlinear distortion estimation in the context of audio.

This paper focuses on the case of memoryless nonlinear distortions, which include common effects such as clipping, rectification, or quantization. This provides an ideal experimental setup for evaluating our method, as these distortions are challenging yet relatively easy to model compared to more complex audio effects. Our goal is to investigate whether the \ac{dps}, previously successful for linear cases, can be applied for nonlinear distortion identification and restoration in the audio domain.
To explore this, a few different models for parameterizing the memoryless nonlinearity are tested.
We demonstrate that the proposed method 
accurately estimates a variety of memoryless nonlinear distortions. Additionally, we evaluate the distortion restoration performance of the proposed method on commonly studied tasks, specifically hard and soft declipping, using both guitar and speech data.

The rest of this paper is organized as follows. Sec.~1 discusses the diffusion model formalism and its use in solving inverse problems in audio. Sec.~2 describes nonlinear function models used in this work, two of which are inherited from previous studies, and the third, the \ac{ccr} 
spline, is tested here for this application for the first time. Sec.~3 presents experimental results. %
Sec.~4 summarizes the research and suggests ideas for future work.

\section{SOLVING BLIND INVERSE PROBLEMS WITH DIFFUSION}
\label{sec:problem}

We consider the following family of inverse problems:
\begin{equation}
    y(n) = f(x(n)), \quad  n \in [0,L-1],
\end{equation}
where $y(n)$ represents the observed distorted measurement, $x(n)$ is an unknown clean audio signal, $n \in \mathbb{Z}$ is the discrete sample index, $L$ is the length of the signal in samples, and $f(\cdot): \mathbb{R} \rightarrow \mathbb{R} $ is an unknown forward operator. 
For convenience in subsequent sections, we introduce vector notation, where the signals are represented as \( \mathbf{y} = [y(0), \dots, y(L)] \in \mathbb{R}^L \) and \( \mathbf{x}_0 = [x(0), \dots, x(L)] \in \mathbb{R}^L \).
The goal is to estimate both the clean audio signal, denoted as $\hat{\mathbf{x}}_0 \approx \mathbf{x}_0$, and the forward operator $f(\cdot)$, given only the distorted measurements $\mathbf{y}$.

Here, $f(\cdot)$ represents a memoryless nonlinear transfer function, meaning its output at any given time index \( n \) depends only on the input at that instant, without influence from past or future inputs.
 We further assume that the distortion introduced by \( f(\cdot) \) is time-invariant over intervals of a few seconds, such that its characteristics remain constant within these segments. Under this assumption, we estimate both \( f(\cdot) \) and \( \mathbf{x}_0 \) using observations from these intervals.

To address the ill-posed nature of this problem, we assume that $\mathbf{x}_0$ is drawn from an underlying data distribution, $\mathbf{x}_0 \sim p_\text{data}$. Additionally, we have access to a training dataset containing realizations from this distribution, where all realizations are normalized to have equal power.
Using this dataset, we train a diffusion model, a type of generative model, that  serves as a data-driven prior of the clean signal distribution. Diffusion models are particularly well-suited for capturing the underlying distribution of complex data, such as audio signals \cite{chen2020wavegrad,
liu2023audioldm, moliner2022solving}, and for generating high-quality samples consistent with this distribution.

Initially, this section offers a summary of diffusion models followed by details on applying them to audio restoration. We further describe the algorithm utilizing diffusion model priors to resolve blind inverse problems.

\subsection{Diffusion Model Formalism}
A diffusion model iteratively transforms Gaussian noise samples $\mathbf{x}_T \sim \mathcal{N}(\mathbf{0}, \sigma^2(T) \mathbf{I})$ to samples drawn from the training data distribution $\mathbf{x}_0 \sim p_\text{data}$.
The generative process is defined in a time interval $\tau \in [0,T] $, with the audio signal at each intermediate state denoted as $\mathbf{x}_\tau \sim \mathcal{N}(\mathbf{x}_0, \sigma^2(\tau) \mathbf{I} )$.  

The generative process can be identified by means of a Probability Flow \ac{ode}: 
\begin{equation}    
    \mathrm{d} \mathbf{x}_\tau=\left[\mathbf{f}\left(\mathbf{x}_\tau, \tau\right)-\frac{1}{2} g^2(\tau) \nabla_{\mathbf{x}_\tau} \log p\left(\mathbf{x}_\tau\right)\right] \mathrm{d} \tau,
\end{equation}
 where time flows backwards from $\tau=T$ to $\tau=0$. The term $\nabla_{\mathbf{x}_\tau} \log p_\tau(\mathbf{x}_\tau)$ is the score function, a vector field that points to areas of higher data likelihood.
The drift $\mathbf{f}(\mathbf{x}_\tau, \tau)$ and diffusion $g( \tau)$ are design parameters.
We adopt the choices from Karras et al. \cite{karras2022elucidating}, which
use $\mathbf{f}(\mathbf{x}_\tau , \tau ) = 0$, $g(\tau) = \sqrt{2\tau}$, and $\sigma(\tau)=\tau$. 
That yields the simplified \ac{ode}:
\begin{equation}\label{eq:ode}
\text{d}\mathbf{x}=  - \tau  \nabla_{\mathbf{x}_\tau} \log p_\tau(\mathbf{x}_\tau) \text{d}\tau.
\end{equation}

The score function $\nabla_{\mathbf{x}_\tau} \log  p_\tau(\mathbf{x}_\tau)$ is intractable in practice, and it thus needs to be approximated.
An intriguing property is that, under Gaussian noise, the score is directly related to an optimal denoiser  $D(\mathbf{x}_\tau, \tau)$ through Tweedie's formula \cite{hyvarinen2005estimation}: 
\begin{equation} \label{eq:tweedie}
    \nabla_{\mathbf{x}_\tau}\log p_\tau(\mathbf{x}_\tau) =(D(\mathbf{x}_\tau,\tau)-\mathbf{x}_\tau)/\sigma^2(\tau).
\end{equation}
That denoiser can be approximated with a deep neural network $D_\theta(\mathbf{x}_\tau, \tau)$ whose weights $\theta$ are optimized through an  $L_2$ denoising objective:
 \begin{equation}\label{loss}
    \mathbb{E}_{\mathbf{x}_0 \sim p_\text{data}, \boldsymbol\epsilon \sim \mathcal{N}(\mathbf{0},\mathbf{I}) } 
    \left[ \lambda(\tau) \lVert D_\theta(\mathbf{x}_0+\tau\mathbf{\boldsymbol\epsilon},\tau) -\mathbf{x}_0   \rVert_2^2 \right],
\end{equation}
where $\lambda(\tau)$ is a time-varying weighting function optimized to ensure the initial loss approximates 1 for all $\tau$, as detailed by Karras et al. \cite{karras2022elucidating}.

\subsection{Diffusion Models for Inverse Problems}\label{sec:dps}
By framing the inverse problem introduced within a probabilistic context, the objective becomes approximating the posterior distribution $ p(\mathbf{x}_0 | \mathbf{y}) $, given the observations $\mathbf{y}$. Solving the \ac{ode} in Eq.~\eqref{eq:ode} from $\tau = T$ to $\tau = 0$ allows for sampling from the learned unconditional distribution. However, in the context of inverse problems, the goal shifts to sampling from the posterior distribution $ p(\mathbf{x}_0 | \mathbf{y})$.

This can be achieved by replacing the score in Eq.~\eqref{eq:ode} by the score of the posterior $\nabla_{\mathbf{x}_\tau} \log p(\mathbf{x} | \mathbf{y})$. Applying Bayes' theorem, this term factorizes as
\begin{equation}
 \nabla_\mathbf{x} \log p(\mathbf{x}|\mathbf{y})=\nabla_\mathbf{x} \log p(\mathbf{x})
+\nabla_\mathbf{x} \log p(\mathbf{y}|\mathbf{x}) .
\end{equation}
 The first term, or \emph{prior score} is approximated with the trained denoiser $D_\theta(\mathbf{x}_\tau, \tau)$ via Eq.~\eqref{eq:tweedie}.
 The second term, or \emph{likelihood score} is intractable in closed form \cite{chung2022diffusion}, and must also be approximated.
 
Following the \ac{dps} framework \cite{chung2022diffusion} and recent extensions \cite{moliner2024buddy,lemercier2024buddyjournal}, we approximate the likelihood score as
\begin{equation}
   \nabla_{\mathbf{x}_\tau} \log p(\mathbf{y}|\mathbf{x}_\tau)  \approx -\zeta(\tau) \nabla_{\mathbf{x}_\tau} C(\mathbf{y}, f(\hat{\mathbf{x}}_0)) ,
\end{equation}
where $C(\cdot, \cdot): \mathbb{R}^L \times \mathbb{R}^L \rightarrow \mathbb{R}$ is a cost function defined as
\begin{equation}\label{eq:cost}
   C(\mathbf{y} , \hat{\mathbf{y}})=
   \frac{1}{M} \sum_{m=1}^M \sum_{k=1}^K
   |S_\text{comp}(\mathbf{y})_{m,k} -S_\text{comp}(\hat{\mathbf{y}})_{m,k} |^2,
\end{equation}
and $S_\text{comp}(\mathbf{y})=|\text{STFT}(\mathbf{y})|^{2/3} \exp{j \angle \text{STFT}(\mathbf{y})}$ is the \ac{stft} spectrogram with compressed magnitudes, $M$ is the number of \ac{stft} frames, and $K$ the number of frequency bins.
The weighting parameter $\zeta(\tau)$ is a time-dependent scaling factor, following \cite{moliner2022solving}:
\begin{equation} \label{eq:variance}
    \zeta(\tau) = \frac{\sqrt{L} \, \tilde{\zeta}}{\sigma(\tau) \|\nabla_{\mathbf{x}} C(\mathbf{y}, f(\hat{\mathbf{x}}_0))\|_2},
\end{equation}
where $\tilde{\zeta}$ is a fixed hyperparameter.

\subsection{Joint Estimation and Restoration Algorithm}

\begin{algorithm}[t]
\caption{Diffusion Inference / operator optimization 
}
\label{alg}

\begin{algorithmic}
\Require distorted measurements $\mathbf{y}$

\State {Sample $\mathbf{x}_{T}\sim \mathcal{N}(\mathbf{y},\sigma^2(T)\mathbf{I})$}
\Comment{Warm initialization}
\State Initialize $\psi_{T}$ 
\Comment{Operator parameters initialization}
\For{$i \leftarrow T, \dots, 1$} \Comment{Discrete step backwards}
    \State $\hat{\mathbf{x}}_0 \leftarrow D_\theta({\mathbf{x}}_{i}, \tau_{i})$ \Comment{Evaluate denoiser model}

    \State $\psi_{i}^0 \leftarrow \psi_{i+1}$  

    \For{$j \leftarrow 0, \dots, N_\text{its.}$}  \Comment{Operator optim. loop}
        \State $\hat{\mathbf{y}} \leftarrow  \hat{f}(\hat{\mathbf{x}}_0;\psi_i^j)$
        \Comment{Evaluate operator}
        \State $\psi_{i}^{j+1} \leftarrow \psi_{i}^{j} - \mathrm{Adam}
        \left( \nabla_\psi C(\mathbf{y},\hat{\mathbf{y}})
        \right)$ \Comment{optim. step}
    \EndFor
    \State $\hat{\mathbf{y}} \leftarrow
    \hat{f}(\hat{\mathbf{x}}_0; \psi_i)$
    \State $\mathbf{l}_\mathbf{x}=- \zeta(\tau_i)\nabla_{\mathbf{x}_i} C(\mathbf{y},\hat{\mathbf{y}}) $ \Comment{Likelihood score estimate}
    \State $\mathbf{s}_i \leftarrow (\mathbf{\hat{x}}_0 - {\mathbf{x}}_i)/\sigma^2(\tau_i)$ 
    \Comment{Tweedie's formula \eqref{eq:tweedie}}
    \State $\mathbf{x}_{{i-1}} \leftarrow {\mathbf{x}}_{i} - {\tau}_i (\tau_{i-1}-{   \tau}_i) (\mathbf{s}_i +\mathbf{l}_\mathbf{x})$ \Comment{Update step}
\EndFor
\State \Return  $\mathbf{x}_0$ , $\psi_0$
\end{algorithmic}
\end{algorithm}

The framework for solving inverse problems introduced in Sec. \ref{sec:dps} typically requires knowledge of the operator $f(\cdot)$. However, in our problem setting, this operator is unknown. To address this, similar to 
 \cite{moliner2023zeroshot,lemercier2024buddyjournal}, we introduce a parametric function $ \hat{f}(\cdot; \psi)$, where $\psi$ is a set of optimizable parameters.
 Practical examples of the studied operators are presented in Sec.~\ref{sec:operators}.

We jointly estimate the clean audio signal $\hat{\mathbf{x}}_0$ and the transfer function parameters $\psi$, as 
proposed recently  \cite{moliner2023zeroshot,lemercier2024buddyjournal}. We aim to minimize the following objective:
\begin{equation}
\arg\min_{\psi, \hat{\mathbf{x}}_0} C(\mathbf{y}, \hat{f}(\hat{\mathbf{x}}_0;\psi)), \quad \text{so that} \quad \hat{\mathbf{x}}_0 \sim p_\text{data},
\end{equation}
where $C(\cdot, \cdot)$ is the cost function defined in Eq. \eqref{eq:cost}.
The constraint $\hat{\mathbf{x}}_0 \sim p_\text{data}$ is weakly enforced by using a diffusion model $D_\theta(\mathbf{x}_\tau, \tau)$ trained using realizations of $p_\text{data}$ as the score-based prior.

The inference algorithm, summarized in Alg. \ref{alg}, implements block-based optimization by alternating state updates of the audio signal $\mathbf{x}_\tau$ with parameter optimization iterations. 
The algorithm combines an Euler sampler for solving the posterior probability-flow ODE introduced earlier in this section.
For optimizing the operator $\hat{f}(\cdot,\psi)$ at a given time-step $\tau$, we first calculate the one-step denoised estimate using $D_\theta(\mathbf{x}_\tau,\tau)$,
then evaluate the cost function, and perform several gradient descent iterations on the parameters $\psi$.
In practice, we employ the Adam optimizer, which accounts for second-order momentum and empirically stabilizes convergence.

Following that procedure, the estimates $\hat{\mathbf{x}}_0$ at the beginning are coarse and contain mostly low-frequency components, 
 leading to suboptimal minimizers for the operator optimization objective.
 Nevertheless, as the process continues, high-frequency details will gradually appear, enabling a more accurate fitting of the operator in the later stages of inference.

In order to accelerate and stabilize inference we employ a warm initialization strategy, similar to \cite{moliner2023zeroshot,lemercier2024buddyjournal}.
This approach involves adding noise to the distorted measurement $\mathbf{y}$ to begin the process, formulated as $\mathbf{x}_T \sim \mathcal{N}(\mathbf{y}, \sigma^2(T)\mathbf{I})$, rather than initializing from pure Gaussian noise.
The algorithm, outlined in Alg. \ref{alg}, is based on a first-order Euler sampler. While we present a deterministic sampler for brevity, the practical implementation incorporates stochasticity following the approach described in \cite{karras2022elucidating}. This stochasticity, which is consistent with prior works \cite{moliner2023zeroshot, lemercier2024buddyjournal}, helps mitigate errors introduced during sampling and provides a modest but significant improvement in results.

\section{NONLINEAR FUNCTION APPROXIMATIONS}\label{sec:operators}

The primary focus of this work is the approximation of an unknown nonlinear degradation operator, defined as $\hat{f}(\cdot; \psi)$.
As explained in Sec.~\ref{sec:problem}, we assume that the degradation is a memoryless nonlinear function. Consequently, we restrict the search space of $\hat{f}(\cdot;\psi)$ to this class of functions.
Importantly, $ \hat{f}(\cdot; \psi)$ must be differentiable with respect to $\psi$ almost everywhere, enabling gradient-based optimization methods such as Adam. This section presents the three nonlinear function models used in the experiments of this study.

\subsection{Sum of Hyperbolic Tangent Functions}

Colonel et al. \cite{colonel2022reverse} proposed a variety of parametric models to reverse engineer nonlinear distortion effects in recording mixes. Unlike their approach, which benefited from having input/output pairs, this work does not have such access, making the task more challenging. After conducting preliminary experiments, we observed that the harmonic sum of hyperbolic tangent functions —SumTanh for short— demonstrated the most stability under these conditions, leading us to concentrate on it.

The mapping using the SumTanh function is defined as 
\begin{equation}
    y(n) = f(x(n)) = \sum_{q=1}^{Q} a(q) \tanh(q\,x(n)),
\end{equation}
where $Q$ is the order of the approximation. This means that a vector~$\mathbf{a}=[a(1),\dots,a(Q)]$ has $Q$ learnable parameters.
Based on our observations we select order $Q=8$.
The composition of only smooth monotonic tanh functions is the main disadvantage of this approach. It does not allow reproducing nonlinear functions with sharp edges or nonmonotonic shapes.

\subsection{Multi-Layer Perceptron}

A \ac{mlp} is a universal function approximator \cite{hornik1989multilayer},  making it a natural choice for modeling nonlinearities.
For example, \ac{mlp}s have been employed as the static nonlinearity component within a Wiener-Hammerstein model to simulate distortion audio effects \cite{kuznetsov2020differentiable}. Furthermore, Hayes et al.~\cite{hayes2021neural} proposed the use of \ac{mlp}s as differentiable waveshapers for sound synthesis.

We use a small four-layer neural network with 20 neurons in both hidden layers. 
Rectified Linear Unit (ReLU) activations, which provide piecewise linear functions, are applied to all layers except the output layer.
The weights are initialized using the Kaiming-normal initialization, while biases are initialized with zeroes. %

The main advantages of \ac{mlp}s are their high expressivity and efficient inference, enabling a large number of iterations to be performed during operator optimization.
However, the network functions as a black box, making it difficult to design appropriate initialization strategies based on the characteristics of the observed distorted data.

\subsection{Cubic Catmull-Rom Splines}

\begin{figure}[t!]
    \centering
    \includegraphics[width=1\linewidth]{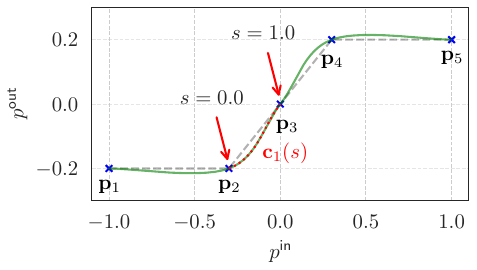}
    \caption{Graphic representation of a \acs{ccr} spline (solid green line) approximating a hard-clipping function (dashed line).
    The red line highlights one curve segment $\mathbf{c}_1(s)$ which is controlled by the points $\mathbf{p}_1$ to $\mathbf{p}_4$.
    }
    \label{fig:ccr}
\end{figure}

Splines are parametric curves that are widely used in computer graphics.
They can fit a wide variety of shapes with a small number of control points \cite{bartels1995}.
One of the most commonly used is a cubic, or third-order spline with four control points and a curve parameter $s \in [0,1)$.
Given $N$ control points, a curve can be constructed by composing partial cubic splines $\mathbf{c}_i$, defined as follows:
\begin{equation}
\label{eq:ccr}
    \mathbf{c}_{ i}(s) = \sum_{j = 0}^{3} b_{j}(s)\mathbf{p} {}_{i + j},\quad i = 0,1, \ldots,N - 3,     
\end{equation}
\noindent
where $b_j$ are elements of a basis vector $\mathbf{b}$,
which defines the shape of the curves 
$\mathbf{c}_i$
between control points $\mathbf{p}_i$ and $\mathbf{p}_{i+1}$.
In particular, the basis vector of a \ac{ccr} spline is defined as~\cite{bartels1995}:
\begin{equation}
\label{eq:basis}
\mathbf{b}(s)=
\begin{aligned} 
    \begin{cases} 
        b_{0}(s) = \frac{1}{2} ( - s + 2s^{2} - s^{3} ), \\
        b_{1}(s) = \frac{1}{2} ( 2 - 5s^{2} + 3s^{3} ), \\
        b_{2}(s) = \frac{1}{2} ( s + 4s^{2} - 3s^{3} ), \\
        b_{3}(s) = \frac{1}{2} ( - s^{2} + s^{3} ).
    \end{cases}
\end{aligned}
\end{equation}
An example of \ac{ccr} is represented in Fig.~\ref{fig:ccr}.
Notably,
as shown in the figure,
the curve $\mathbf{c}_1(s)$ connects two of the inner control points $\mathbf{p}_2$ and $\mathbf{p}_3$.
This property allows accurate control across points, which facilitates initialization strategies.

To describe the distortion operator $\hat{f}(\cdot, \psi)$, we use $N + 2$ control points
$ \mathbf{p}_{0}, \dots \mathbf{p}_{N+1} $.
Each control point 
$ \mathbf{p}_{i}=[p^{\text{in}}_i, p^{\text{out}}_i]$ is composed of a 
grid point
(input value) $p^{\text{in}}_i$ and an output value $p^{\text{out}}_i$.
The set of output values are the optimizable parameters 
$\psi=\{p_0^\text{out},\dots,p_{N+1}^{\text{out}}\}$.
\newcommand{\pin}{p^{\text{in}}}
The grid points $p^{\text{in}}_1, \dots, p^{\text{in}}_N$ are fixed and defined in the range $p_i^\text{in} \in (-1,1)$.
After initializing the grid points $p_i^{\text{in}}$ uniformly in the range $(-1,1)$, we apply the following transformation to each point to map it according to the $\mu$-law:
\begin{equation}
p^{\text{in}}_{i} \leftarrow     
\mathrm{sgn}(p^{\text{in}}_i) 
\dfrac{(1+\mu)^{\lvert{p^{\text{in}}_i}\rvert}}{\mu} .
\end{equation}
This transformation increases the density of control points near zero with higher $\mu$, aligning with the natural clustering of audio data, which often exhibits a higher concentration of values close to zero.
The extremal points are set to $p^{\text{in}}_0 < -1 $ and $ p^{\text{in}}_{N+1} > 1$ to ensure control over the curve throughout the whole range $(-1, 1)$.
An odd number of control points is used to have one point at zero.
In our experiments, we use $N=41$ points and $\mu=20$.

To evaluate the distortion operator $\hat{f}(\cdot, \psi)$ during runtime, the input values $ x(n) $ are discretized into bins defined by the grid points $ p_i^{\text{in}}$. For each bin, the curve parameter $s \in [0,1)$ is calculated based on the relative position of $x(n)$ within the interval. The corresponding spline segment $\mathbf{c}_i(s)$ is then evaluated by determining the basis vector $\mathbf{b}(s)$ and applying Eq. \eqref{eq:ccr}. This operation is fully differentiable with respect to the parameters $\psi$, allowing for gradient-based optimization.

\section{EXPERIMENTS AND RESULTS}

Since two tasks are solved in parallel in this study, experiments and results are presented in separate parts as well. The first part, Sec. \ref{sec:operator_estimation}, discusses how accurate the method is in estimating the unseen degradation operator. The second, Sec. \ref{sec:guitar_restoration} compares, the proposed method for generating plausible restored guitar audio signals.
Finally, Sec. \ref{sec:speech_restoration} presents results on speech restoration.

The proposed method is designed to restore audio signals affected by any memoryless nonlinear distortion. However, for the experiments in this study, we focus specifically on the restoration of signals distorted by hard and soft clipping. This choice is motivated by their practical relevance and broad applicability, as well as the availability of well-established baselines for comparison.

\subsection{Operator Estimation}\label{sec:operator_estimation}

\begin{figure*}[t]
    \centering
    \includegraphics[width=1\linewidth]{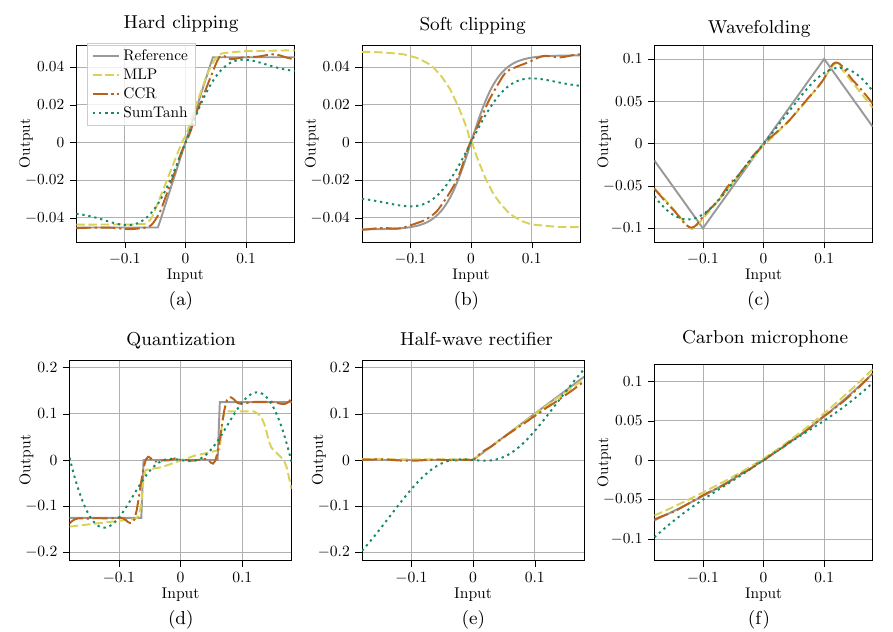}
    \caption{Memoryless nonlinearities and the corresponding blind estimates with the three different models. The proposed method CCR is the most accurate one in most cases.
    }
    \label{fig:operator_approx}
\end{figure*}

The first experiment evaluates how effectively the proposed approach generalizes across various memoryless nonlinear distortions and compares the performance of the nonlinear function approximations introduced in Sec. \ref{sec:operators}.

\subsubsection{Nonlinear Distortions Evaluated}

Figure \ref{fig:operator_approx} depicts the ramp responses of the evaluated nonlinear functions, along with their blind estimates obtained using the different waveshaper methods. The specific nonlinear distortions studied are as follows:
\begin{enumerate}[label=\alph*)]
     \item Hard clipping: A classic distortion that applies a threshold, resulting in a sharply truncated waveform. In this study, we evaluate hard clipping with an input SDR of 3 dB.
     \item Soft clipping: Similar to hard clipping but with smooth transitions at the thresholds, making it less trivial to identify. We evaluate soft clipping with the same input SDR of 3 dB.
     \item Wavefolding: A non-monotonic function where portions of a waveform are ``folded back'' on itself \cite{esqueda2017virtual}.
     \item Quantization: 
      This distortion occurs when signal values are quantized to a finite number of levels \cite{zaviska2021,mokry2020flexible}.
      We study an extreme case with only three quantization levels that can be achieved with two bits. 
     \item \ac{hwr}: An asymmetric nonlinearity where only the positive half of the signal is preserved, while the negative values are set to zero. 
 \end{enumerate}

Additionally, although not included in the evaluation, we experiment with a carbon microphone distortion, a nonlinearity characteristic of early recording technologies, which introduces asymmetric waveshaping \cite{Quatieri2000, oksanen2011carbon}. This distortion is represented in Fig. \ref{fig:operator_approx}f.
All details and audio examples are available on the project website \footnote{\href{https://michalsvento.github.io/NLDistortionDiff/}{https://michalsvento.github.io/NLDistortionDiff/}}.

\subsubsection{Experimental Details}

We train a diffusion model on Direct Input clean guitar data.
Our study utilizes the IDMT-SMT-GUITAR dataset \cite{kehling2014automatic}, specifically its fourth subset, which includes 64 brief musical pieces performed at varying tempos. These compositions were captured using two distinct electric guitars; however, we exclusively use the Career SG guitar.
We allocated 5\,min for testing, 5\,min for validation, and the remaining 26.5\,min for training. The evaluation dataset was split into 50 segments, each approximately 6-s long, which were employed for objective evaluation.
The data from this experiment was sampled at 44.1\,kHz.

Given the limited volume of data available for training, it was observed that exclusive training of the diffusion model on this dataset resulted in significant overfitting, evidenced by the model's tendency to memorize the data.
In order to minimize this issue, we opted for pre-training the diffusion model with a related larger dataset, and then fine-tuned the model for only a small number of iterations. 
Specifically, we use the Direct Input recordings from the EGDB dataset \cite{chen2022towards}, which contains around 2\,h of audio.

The model architecture is based on a CQT-based U-Net from \cite{moliner2024diffusion}, with approximately 40 million parameters. We used the AdamW optimizer with weight decay (0.01), a learning rate of 2e-5, and a batch size of 4 segments, each 6\,s long. The training process involved 290\,000 iterations of pre-training followed by 26\,000 iterations of fine-tuning. Exponential Moving Average (EMA) weights were tracked with a rate of 0.9999 and applied during evaluation.

For inference, we performed $T=50$ steps of reverse diffusion with a \ac{dps} step size $\zeta^\prime=0.3$ and 20 optimization steps for operator estimation, using a learning rate of 0.02. For additional details, we refer to the source code \footnote{
\href{https://github.com/michalsvento/NLDistortionDiff}{https://github.com/michalsvento/NLDistortionDiff}
}. 
We aimed to ensure a fair comparison by selecting hyperparameters that worked well across all of the nonlinear function approximation models.

\subsubsection{Evaluation Metrics and Results}

In our evaluation, we use \ac{lsd} and \ac{rrmse}.
Given clean guitar test samples $\mathbf{x}_0$, \ac{lsd} quantifies the similarity between the blindly predicted distorted signal $\hat{f}(\mathbf{x}_0, \psi)$
and the reference distorted signal $f(\mathbf{x}_0)$ in the log-spectral domain.
Given $Y \in \mathbb{C}^{M \times K}$ the \ac{stft} of the reference distorted audio and $\hat{Y} \in \mathbb{C}^{M\times K}$ the \ac{stft} of the predicted one,
we compute \ac{lsd} as follows:
\begin{equation}
\begin{split}
    \text{LSD} &= \frac{1}{M \times K} 
    \sum_{m=1,k=1}^{M,K}
    \big(
    \log_{10} (|Y_{m,k}|^2 + \varepsilon) \\
    &\hspace{3cm}
    - \log_{10}(|\hat{Y}_{m,k}|^2 + \varepsilon)
    \big)^2
\end{split}
\label{eq:log_spectral_distortion}
\end{equation}

To calculate \ac{rrmse}, we use as input a ramp signal $\textbf{r}\in \mathbb{R}^P$ with $P=1000$ linearly-spaced points in the range
$[-3\sigma_{\textrm{data}},3\sigma_{\textrm{data}}]$,
where $\sigma_\text{data}$ represents the standard deviation of the clean data, computed from the  dataset.
The metric is defined as: 
\begin{equation}
    \textrm{RR-MSE}
    =
    \dfrac{1}{P}\sum_{p=1}^{P}
    (
    f(\mathbf{r}_p)- \hat{f}(\mathbf{r}_p,\psi)
    )^2.
    \label{eq:MSEloss}
\end{equation}
In some cases, we observed that the operator estimations converge to a result with a sign flip (see Fig. \ref{fig:operator_approx}b for an example). While this may seem unexpected, we argue that this phenomenon should not be penalized, as it does not affect the audio quality.
This sign flip is an expected consequence of the blind estimation paradigm, where an audio signal estimate with a flipped sign is equally likely according to the probability distribution inferred from the diffusion model. Although this does not impact the \ac{lsd}, it does significantly influence \ac{rrmse}.
To account for this, we also evaluate \ac{rrmse} using a sign-flipped operator and select the result that minimizes the error.

Table \ref{tab:operator_eval} presents the average results of the experiment across the different evaluated nonlinearities. 
The \ac{ccr} consistently outperforms the other methods in nearly all cases, except for the quantization scenario, in which the \ac{mlp} achieves a lower \ac{lsd}, indicating superior performance in~this specific instance. However, this trend does not hold for the \ac{rrmse} case, where \ac{ccr} maintained its advantage.

Figure \ref{fig:operator_approx} exemplifies blind transfer function estimates generated by the three nonlinear function models. A single randomly selected estimate is shown for each example, which may not fully represent the performance across the entire dataset. The figures show that the SumTanh model struggles to fit asymmetric nonlinearities, as observed in the \ac{hwr} case (Fig. \ref{fig:operator_approx}e). Additionally, the SumTanh model has difficulty handling sharp transitions, making it most effective for soft clipping (Fig. \ref{fig:operator_approx}b), compared to the others. The \ac{mlp}, however, is good at fitting sharp edges and may outperform CCR in this regard, as CCR tends to introduce small errors at these transitions (see Figs. \ref{fig:operator_approx}a and \ref{fig:operator_approx}b). Nevertheless, CCR  generally provides more robust estimates than \ac{mlp}, as indicated by the metrics given in Table \ref{tab:operator_eval}. We speculate that the advantage of CCR lies in its optimization stability, partly due to easy initialization.

\begin{table}[th]
\centering
\caption{Comparison of three nonlinear function models in~approximating five nonlinearities.
The metrics \ac{lsd} and \ac{rrmse}
are reported in dB scale. 
The best results are bolded.}
\label{tab:operator_eval}
\adjustbox{width=\columnwidth}
{
\begin{tabular}{@{}llrr@{}}
\toprule
\multicolumn{1}{l|}{Nonlinearity} & Method  &  LSD (dB)$\downarrow$ & RR-MSE (dB)$\downarrow$ \\ \midrule
\multicolumn{1}{l|}{Hard clipping}    & SumTanh & 15.19 & {-46.25}\\
\multicolumn{1}{l|}{}      & MLP     & 3.77  & -47.32 \\
\multicolumn{1}{l|}{}      & CCR     & \textbf{2.51}  &\textbf{-54.82} \\ \midrule
\multicolumn{1}{l|}{Soft clipping}    & SumTanh & 3.27  & {-49.67} \\
\multicolumn{1}{l|}{}      & MLP     & 6.71  & -47.27 \\
\multicolumn{1}{l|}{}      & CCR     & \textbf{5.94}  & \textbf{-56.81} \\ \midrule
\multicolumn{1}{l|}{Wavefolding}    & SumTanh & 13.41 & {-36.57}\\
\multicolumn{1}{l|}{}      & MLP     & 4.09  & -35.60 \\
\multicolumn{1}{l|}{}      & CCR     & \textbf{3.74}  & \textbf{-39.29} \\ \midrule
\multicolumn{1}{l|}{Half-wave}   & SumTanh & 13.00 & -23.98 \\
\multicolumn{1}{l|}{rectification}      & MLP     & 4.07  &-39.45 \\
\multicolumn{1}{l|}{}      & CCR     &\textbf{ 2.94}  &\textbf{ -50.61} \\ \midrule
\multicolumn{1}{l|}{Quantization} & SumTanh & 68.66 & {-27.33}\\
\multicolumn{1}{l|}{}      & MLP     & \textbf{4.72} & -32.62 \\
\multicolumn{1}{l|}{}      & CCR     & 15.64 &\textbf{ -35.86} \\ \bottomrule
\end{tabular}
}
\end{table} 

\subsection{Distorted Guitar Restoration}\label{sec:guitar_restoration}

We evaluate the restoration of guitar signals, focusing on hard and soft declipping. The same dataset and implementation details as in Sec.~\ref{sec:operator_estimation} are used for consistency.
We analyze three degrees of distortion: input \ac{sdr}s of 1-dB, 3-dB, and 7-dB, where lower means higher distortion.

\subsubsection{Compared Methods}

As a baseline for guitar restoration, we used the Demucs network\footnote{\href{https://github.com/facebookresearch/denoiser}{https://github.com/facebookresearch/denoiser}}, trained in a supervised manner over 100\,000 iterations, following the approach outlined in \cite{Imort2022distortion}. We trained specialized models separately for hard and soft clipping tasks to address their distinct characteristics.

For the hard clipping task, the input signal-to-distortion ratio (SDR) was sampled from the range $[0.1, 10]$, and the clipping threshold was adjusted accordingly. For the soft clipping task, the model was trained on the same SDR range. However, the degradation operator for soft clipping was parameterized as described in \cite{defreaene2012perceptclip}. This parameterization employs an adjustable hardness level $r \in [0,1]$, sampled uniformly at random during training.
At the extremes, when $r=1$, the distortion closely resembles hard clipping, whereas $r=0$ produces a smooth, $\mathrm{tanh}$-like degradation.
This setup aims to model a sufficiently diverse range of soft clipping distortions.

We also include, as a baseline, an informed version of our proposed algorithm. In this setup, we utilize the ground-truth degradation operator $f(\cdot)$ instead of blindly optimizing it. This corresponds to the method described in Sec. \ref{sec:dps} and previously investigated for hard clipping in prior works \cite{moliner2022solving, hernandez2024vrdmg}.

\subsubsection{Evaluation Metrics and Results}

\begin{figure*}[t]
    \includegraphics[]{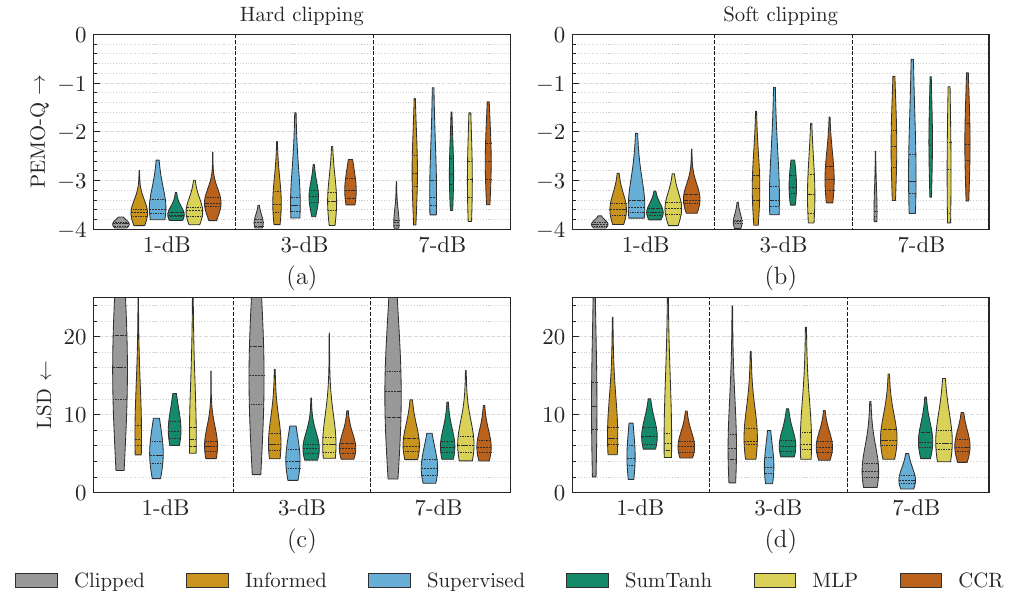}
    \caption{Violin plots of PEMO-Q and LSD results evaluated on 50 selected samples from IDMT-SMT-GUITAR test set. Every bin represents defined input \ac{sdr}.}
    \label{fig:guitar_pemoq_lsd}
\end{figure*}

We report two reference-based metrics: \ac{lsd} and PEMO-Q.
\ac{lsd} quantifies the spectral distortion between the restored signal and the target. PEMO-Q is designed to estimate the perceptual quality of audio, focusing on aspects that are important for human perception.
The results, summarized in Fig. \ref{fig:guitar_pemoq_lsd}, are presented as violin plots for both metrics.
The supervised baseline outperforms the proposed methods in terms of LSD. However, some of the proposed methods achieve better performance on PEMO-Q. This aligns with expectations, as LSD is not well-suited for evaluating generative models, whereas PEMO-Q, being a perceptual metric, provides a more relevant assessment.
In the 1-dB \ac{sdr} case for both hard and soft clipping, the \ac{ccr} method demonstrates superior performance compared to all other approaches. For the 3-dB and 7-dB \ac{sdr} cases, \ac{ccr} shows marginally better results.
Interestingly, \ac{ccr} surpasses the informed setting in certain scenarios, particularly for the 1-dB case (see Figs.~\ref{fig:guitar_pemoq_lsd}a and b). 

In addition, we report the \ac{fad}~\cite{roblek2019fr},
which quantifies the similarity between the distributions of real and generated audio embeddings.
\ac{fad} is computed by comparing embeddings of the original and restored signals. Using the fadtk tool \cite{gui2024adapting}, we compute \ac{fad} with both the original VGGish embeddings and the embeddings from the audio compression model EnCodec \cite{defossez2022high}. \ac{fad} evaluates the quality of the generated audio in terms of both timbral and temporal aspects but does not directly capture hallucinations or instances where the restored audio may not match the content of the distorted measurement.

\begin{figure}[t]
    \centering
    \includegraphics[width=1\linewidth]{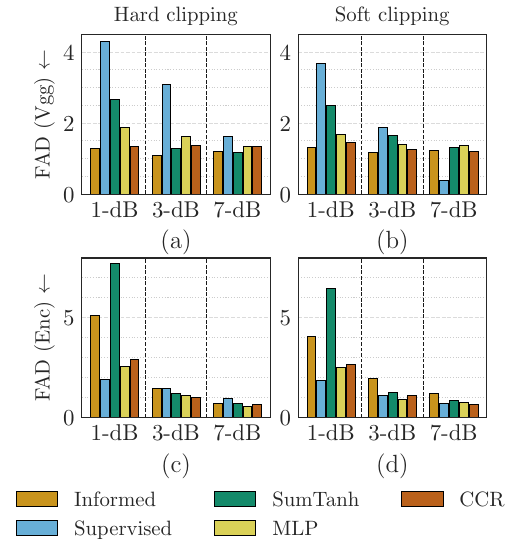}
    \caption{Results from \ac{fad} using  embeddings from Vggish (Vgg) and EnCodec (Enc) on the IDMT-SMT-GUITAR test set.}
    \label{fig:guitar_fad}
\end{figure}

The results for \ac{fad} are shown in Fig. \ref{fig:guitar_fad}.
When computed using VGGish embeddings,
FAD heavily penalizes the supervised baseline at 1-dB and 3-dB \ac{sdr}. 
Most methods achieve comparable scores for the remaining conditions,
including the informed setting, except SumTanh,
which performs worse at 1-dB \ac{sdr}.
When computed with EnCodec embeddings, the supervised baseline is less penalized and even achieves the best performance at 1-dB \ac{sdr}. However, it is slightly outperformed by the blind methods in hard clipping scenarios at 3-dB and 7-dB \ac{sdr}. For soft clipping, the scores for the supervised baseline and \ac{ccr} are comparable. Interestingly, in the EnCodec-based evaluation, the informed setting performs worse than both the supervised and blind methods for both hard and soft clipping at 1-dB and 3-dB \ac{sdr}.

\vspace{-0.2cm}
\subsection{Speech Restoration}\label{sec:speech_restoration}

In this section, we evaluate the proposed method on speech hard/soft declipping, a task with established relevance.
A comparison with recent baselines is presented.

\vspace{-0.2cm}
\subsubsection{Experimental Details}

We experiment with the VCTK dataset \cite{valentini2016reverberant}, which contains studio-quality speech recordings.
We employ the same train split as Yi et al.~\cite{yi2024ddd}, to facilitate the comparison. 
Additionally, the dataset is resampled to 16 kHz. 
The evaluation set contains 100 utterances from the two speakers in the test split:``p232'' and ``p257''.

We use an STFT-based NCSN++M architecture that has proven successful for speech diffusion models \cite{richter2022speech, moliner2024buddy}, whereas the original NCSN++ was proposed for image generation \cite{song2020denoising}.
The model contains 27.8 million parameters.
The training setup is similar to that described in Sec.~\ref{fig:operator_approx} for the guitar, with a batch size of 8. The model undergoes a single-stage training process without finetuning for approximately 300\,000 iterations.
The inference details remain similar, but we opted for $T=100$ sampling steps. For further specifications, please refer to the source code \footnote{\href{https://github.com/michalsvento/NLDistortionDiff}{https://github.com/michalsvento/NLDistortionDiff}}.

\subsubsection{Compared Methods and Baselines}

We evaluate the proposed method for speech hard and soft declipping using the \ac{mlp} and \ac{ccr} models, comparing it against several representative baselines. The baselines include A-SPADE \cite{kitic2015sparsity}, a sparsity-based approach designed for hard clipping restoration, and DDD \cite{yi2024ddd}, a supervised model specifically trained for hard clipping scenarios. Additionally, we consider general-purpose speech enhancement tools, such as VoiceFixer \cite{liu2022voicefixer}, an open-source model for audio restoration, and Resemble Enhance~\footnote{\href{https://github.com/resemble-ai/resemble-enhance}{https://github.com/resemble-ai/resemble-enhance}},
another recent speech enhancement tool.
Finally,
we include the Informed setting,
where the conditioning provides knowledge of the true clipping function.

\subsubsection{Evaluation Metrics and Results}

We evaluate the performance of the proposed method using two metrics. NISQA \cite{mittag2021nisqa} is a non-intrusive metric that estimates speech quality without requiring a reference signal. ESTOI \cite{jensen2016algorithm}, on the other hand, is a reference-based metric designed to evaluate speech intelligibility.  

\begin{figure*}[t]
    \includegraphics[]{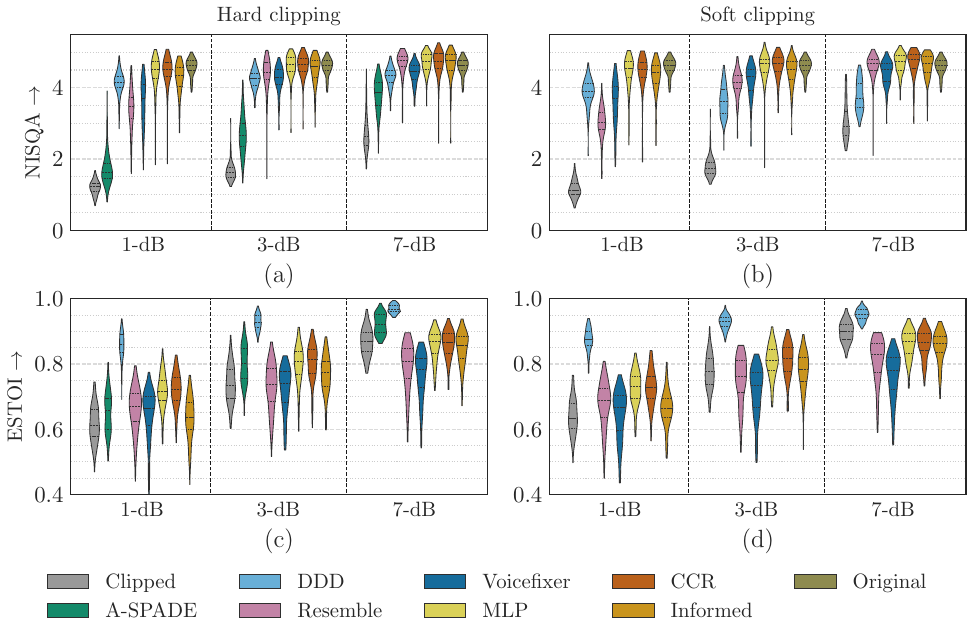}
        \caption{Violin plots of NISQA and ESTOI results evaluated on 100 selected samples from the VCTK test set. Every bin represents defined input \ac{sdr}.}
    \label{fig:speech_violin}
\end{figure*}

The results are presented as violin plots in Fig. \ref{fig:speech_violin}. For NISQA, the proposed blind methods consistently achieve the best performance and, in some cases, even surpass the scores of the original signal (Fig. \ref{fig:speech_violin}a and Fig. \ref{fig:speech_violin}b). The informed method performs slightly worse than the blind approaches, although the differences are not always statistically significant.  

For ESTOI, the supervised baseline DDD significantly outperforms all other methods (Fig. \ref{fig:speech_violin}c and Fig. \ref{fig:speech_violin}d).
Interestingly, A-SPADE and the clipped condition,
which performed poorly on NISQA,
achieve relatively high scores on ESTOI.
 We attribute this to phase differences that, while minimally perceptible, can significantly influence the ESTOI metric.
Notably, the proposed blind methods surpass the speech enhancement baselines, VoiceFixer and Resemble Enhance.
Overall, the results demonstrate that the proposed method achieves state-of-the-art performance for blind distorted speech restoration, delivering a strong balance between speech quality and intelligibility.

\vspace{-0.2cm}
\section{CONCLUSION}

We introduced a novel approach for the blind estimation and inversion of memoryless nonlinearities in music and speech recordings. This method represents the first general application of diffusion models and cubic Catmull-Rom splines for modeling nonlinear audio distortions. The approach operates in challenging scenarios where both the original input signal and the applied nonlinearity are unknown.
Additionally, the method is unsupervised, utilizing only distorted measurements and a diffusion model that is trained on clean audio files.

The proposed method demonstrated strong performance in estimating a diverse set of memoryless nonlinearities and restoring clipped music and speech signals. 
Among the nonlinear function approximation models, \ac{ccr} splines emerged as the most effective overall.
While in the task of audio declipping, the method underperformed supervised baselines in reference-based metrics, it excelled in non-intrusive metrics such as NISQA, FAD, and the perceptually motivated PEMO-Q metric. 

Memoryless systems, while limited in scope, provide a valuable academic exercise for understanding and addressing audio distortions in difficult blind and unsupervised settings. 
However, in real-world applications, distortion effects are typically more complex and often involve memory. Previous work has demonstrated promising results in modeling linear systems using similar approaches \cite{moliner2023zeroshot, lemercier2024buddyjournal}, but not with the same level of accuracy achieved for memoryless distortions in this study.
Future work should focus on extending this method to address more complex nonlinear effects with intricate dynamics.

\vspace{-0.5cm}
\section{ACKNOWLEDGMENT}
\vspace{-0.2cm}
This study was conducted during a four-month Erasmus+ traineeship of the first author at the Aalto Acoustics Lab in August--December 2024. The work of the first author was also supported by the Czech Science Foundation (GA\v{C}R) Project No.\,23-07294S. The authors acknowledge the computational resources provided by the Aalto Science-IT project. The authors are grateful to Prof.~Pavel Rajmic for his support and helpful discussions.

\bibliography{jaes.bib}
\bibliographystyle{jaes.bst}

\appendix

\biography{Michal Švento}{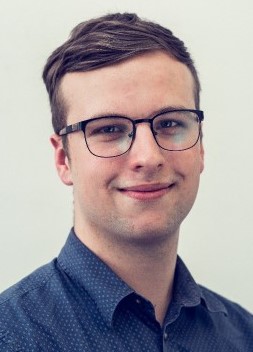}{
Michal Švento completed his B.Sc. and M.Sc. degrees in Audio Engineering at Brno University of Technology, Czechia, in 2021 and 2023.
He is currently a doctoral student in the Telecommunications program at the university's Department of Telecommunications.
His research focuses on solving audio inverse problems by combining machine learning methods with traditional approaches.
}

 \biography{Eloi Moliner}{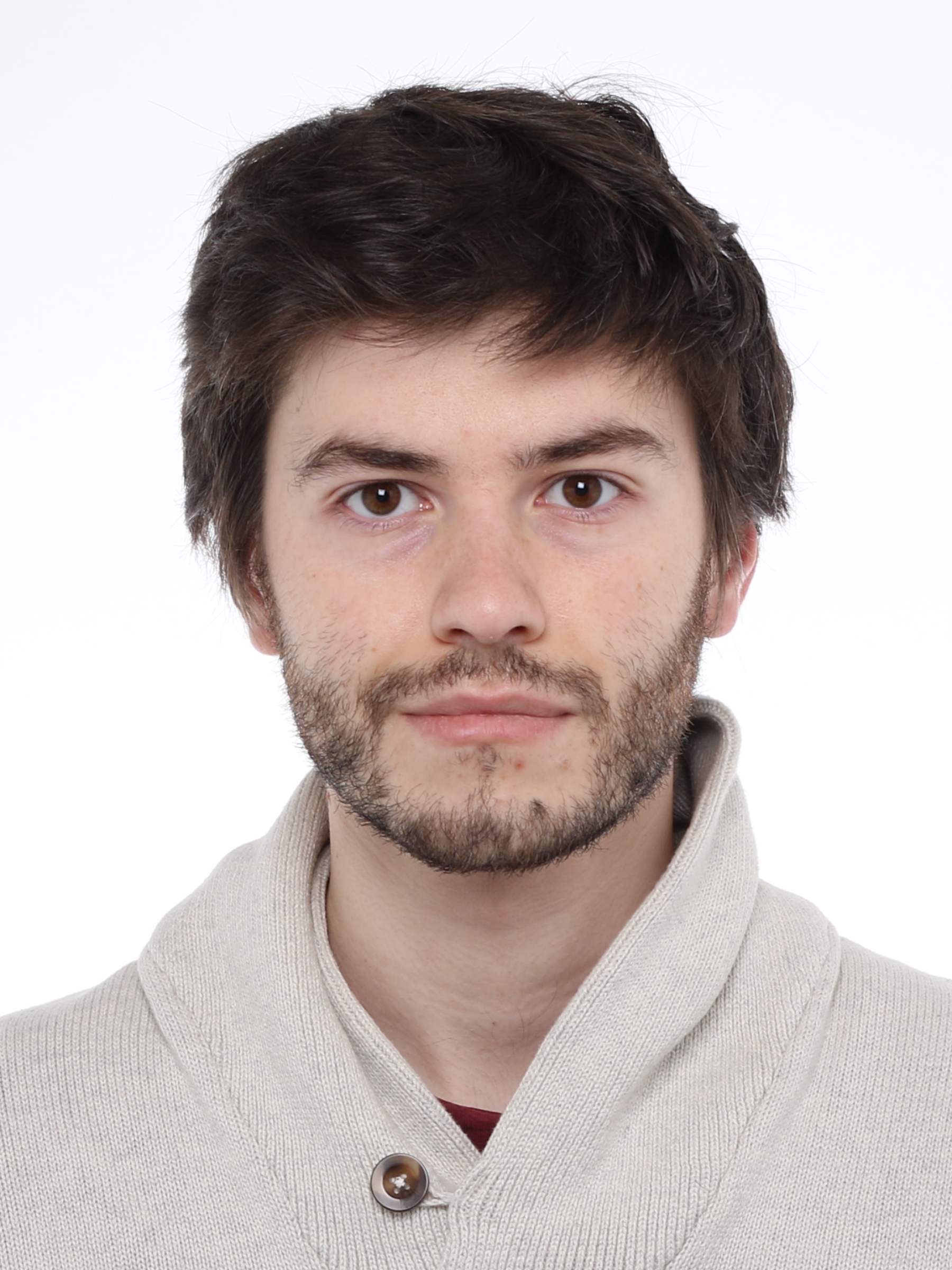}{Eloi Moliner received his B.Sc. degree in Telecommunications Technologies and Services Engineering from the Polytechnic University of Catalonia, Spain, in 2018 and his M.Sc. degree in Telecommunications Engineering from the same university in 2021.
He is currently a doctoral candidate at the Acoustics Lab of Aalto University in Espoo, Finland. His research interests include digital audio restoration and audio applications of machine learning. He is the recipient of the Best Student Paper Awards at the 2023 IEEE ICASSP conference and at the 2024 IWAENC conference.}

\biography{Lauri Juvela}{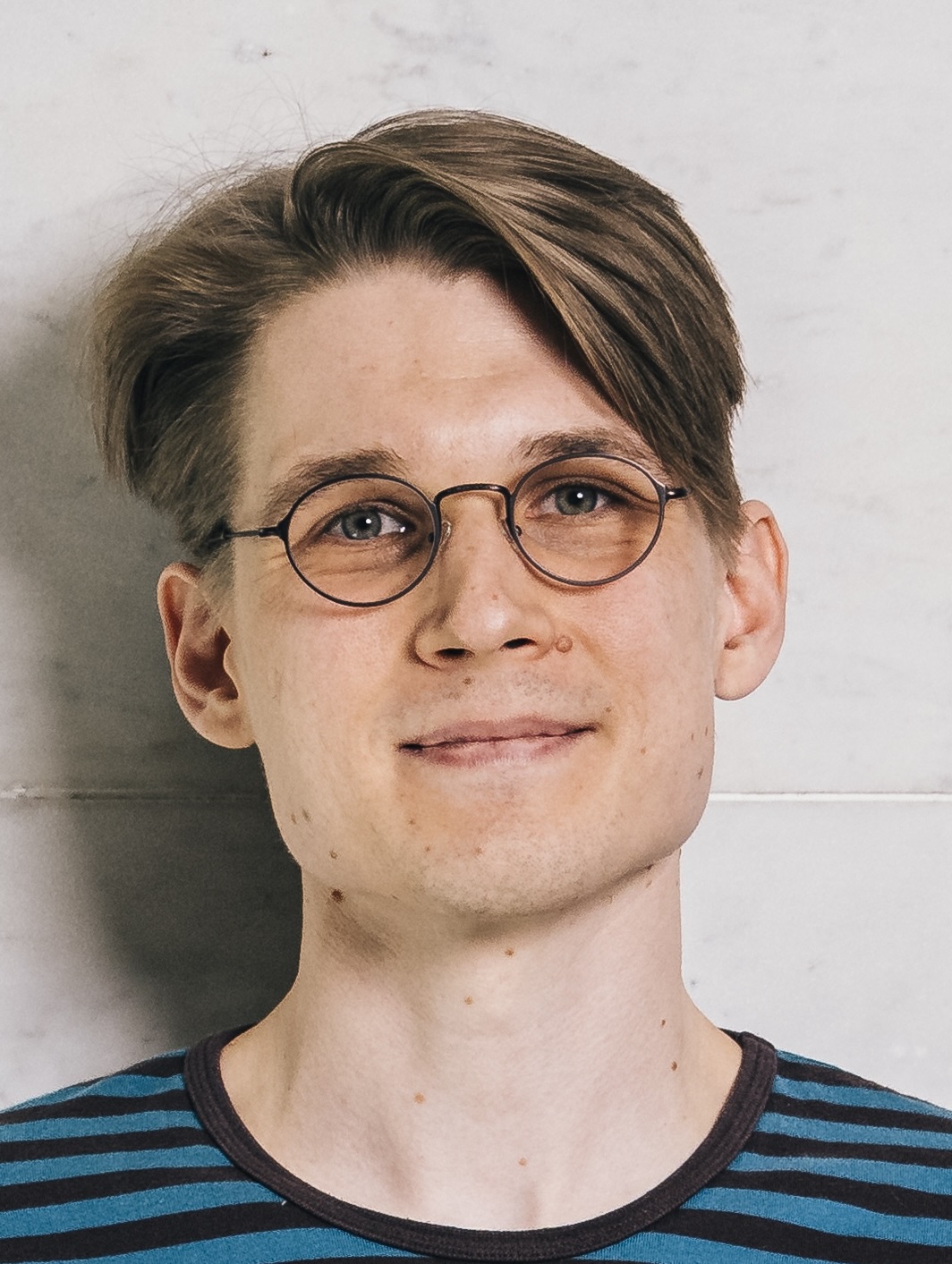}{Lauri Juvela is an Assistant Professor in Machine Learning for Speech and Language Technology at Aalto University, Finland. He earned his D.Sc. degree on speech synthesis at Aalto University in 2020 and worked as a Machine Learning Researcher at Neural DSP Technologies in 2019-2023. His research interests include deep generative models for speech and audio.}

\biography{Alec Wright}{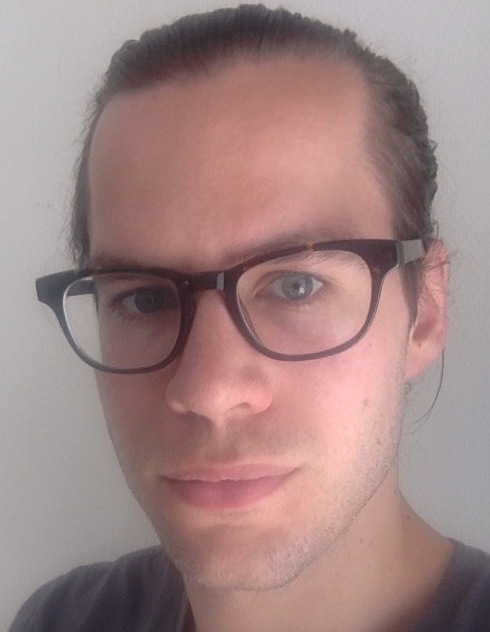}{Alec Wright is a Chancellor's Fellow in Audio Machine Learning at the University of Edinburgh. He received his D.Sc degree from Aalto University, Finland, in 2023. His research interests include virtual analog modeling of guitar amplifiers and other audio effects, and machine learning for audio applications.}

\biography{Vesa Välimäki}{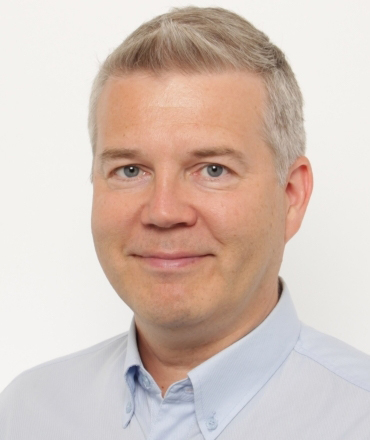}{Vesa V\"alim\"aki is Full Professor of audio signal processing and Vice Dean for Research at Aalto University, Espoo, Finland. He received his D.Sc. degree from the Helsinki University of Technology in 1995. In 1996, he was a Postdoctoral Research Fellow at the University of Westminster, London, UK. In 2008--2009, he was a visiting scholar at  Stanford University. 
He is a Fellow of the AES, the IEEE, and the Asia-Pacific Artificial Intelligence Association. Prof.~V\"alim\"aki is the Editor-in-Chief of the Journal of the Audio Engineering Society.}

\end{document}